# Noninvasive in vivo photoacoustic measurement of internal jugular venous oxygenation in humans

Alejandro Garcia-Uribe[1], Todd N. Erpelding[2], Haixin Ke[1], Kavya Narayana Reddy[3], Anshuman Sharma[3], and Lihong V. Wang[1,4]

[1]*Optical Imaging Laboratory, Department of Biomedical Engineering, Washington University in St. Louis, St. Louis, Missouri, USA*

[2]*Philips Research North America, Briarcliff Manor, NY, USA*

[3]*Department of Anesthesiology, Washington University School of Medicine, St. Louis, MO, USA*

[4]*Current address: Caltech Optical Imaging Laboratory, Andrew and Peggy Cherng Department of Medical Engineering, Department of Electrical Engineering, California Institute of Technology, 1200 E. California Blvd., MC 138-78, Pasadena, CA 91125, USA; lvw@caltech.edu*

# ABSTRACT

In many clinical conditions, such as head trauma, stroke, and low cardiac output states, the brain is at risk for hypoxic-ischemic injury. The metabolic rate and oxygen consumption of the brain are reflected in internal jugular venous oxygen saturation ($s_{ijv}O_2$). The current gold standard for monitoring brain oxygenation is invasive; it requires jugular vein catheterization under fluoroscopic guidance and therefore is rarely used. Photoacoustic (PA) measurement, on the other hand, can estimate the oxygen consumption of the brain non-invasively in real time. Such a convenient method can potentially aid earlier detection and prevention of impending hypoxic brain injury.

A dual-wavelength photoacoustic tomography (PAT) and ultrasound imaging (US) system was used to image the internal jugular vein (IJV) and estimate the $s_{ijv}O_2$ in seven healthy volunteers. The system captured simultaneous co-registered PAT and US images at a rate of five frames per second. For each volunteer, the internal jugular vein was identified under ultrasound guidance, then, additional PA images from two optical wavelengths were collected and used to estimate the oxygen saturation of the internal jugular vein.

For each volunteer, the oxygen saturation was calculated from transverse and longitudinal views of the internal jugular vein. The average $s_{ijv}O_2$ was $72 \pm 7$ %. The preliminary results are encouraging and agree with those reported in the literature.

Photoacoustic images were successfully used to calculate the blood hemoglobin oxygen saturation in the internal jugular vein. These results raise confidence that this emerging technology can be used clinically for accurate, noninvasive estimation of $s_{ijv}O_2$. In addition, the fast co-registration with US images makes the technique suitable for studying the temporal variations of oxygen saturation in response to physiologic challenges in clinical settings.

# INTRODUCTION

Lessons learned from epidemiological efforts in the area of traumatic brain injury suggest that even brief episodes of hypoxia are associated with significant increases in mortality and morbidity [1]. Similarly, brain hypoxia/ischemia can also occur during many complex surgical or diagnostic procedures, such as carotid endartrectomy, cardiac surgeries, and neuro-interventional radiological procedures. In addition, patients admitted to critical care units for reasons such as traumatic brain injury, multi-organ failure, septicemia, and prematurity, frequently develop cardio-respiratory insufficiency, which often leads to brain ischemia.

Even with extensive monitoring of hemodynamic parameters like arterial blood pressure and arterial oxygen saturation, central venous pressure imbalance in the oxygen supply and metabolic demands of brain tissue can persist and go undetected. Improving oxygen delivery to match metabolic demands is essential for managing patients whose brain tissue is anticipated to be at risk from ischemic injury. Monitoring of mixed venous oxygen saturation has been shown to detect inadequate systemic oxygen delivery in patients with normal cardiac output and arterial oxygen saturation. Goal-directed therapy based on mixed venous oxygen saturation has been shown to improve survival and reduce the incidence of postoperative organ failure in critically ill adults and children.

Monitoring oxygen saturation in the venous blood draining from the brain is one reliable way of monitoring the adequacy of global brain perfusion, and it is frequently used in patients with traumatic brain injury. Changes in global oxygen consumption are closely related to changes in $s_{ijv}O_2$. All these advantages make $s_{ijv}O_2$ almost an ideal marker for global cerebral perfusion. Unfortunately, monitoring $s_{ijv}O_2$ requires placing a vascular catheter in the jugular bulb, and thus can be performed only in the operating room or under fluoroscopic guidance. Routine use of $s_{ijv}O_2$ is further limited because of many other challenges, such as the development of micro-thrombi at the tip of the catheter and the need for frequent re-calibration. In addition, one has to predict the dominant side of the venous drainage prior to the placement of the invasive catheter, making routine monitoring of $s_{ijv}O_2$ impractical.

In this work, we demonstrate the feasibility of a novel non-invasive method that has the potential for bedside monitoring of internal jugular vein oxygen saturation. This method combines principles of photoacoustics and ultrasonic imaging and can be used to measure $s_{ijv}O_2$ in real time without the need for direct vascular access. Because changes in global oxygen consumption are closely reflected in changes of $s_{ijv}O_2$ values, we believe that the bedside ability to measure and monitor $s_{ijv}O_2$ will offer a major advantage in the management of critically ill patients.

## MATERIALS AND METHODS

### 2.1 Fundamentals of Photoacoustic Tomography

PAT of internal organs has been investigated in animal models for several years [2, 3]. PAT can offer structural, functional, and molecular contrasts from typical endogenous tissue chromophores (optical absorbers), including hemoglobin, melanin, lipid, and water. The optical absorption of these absorbers is sensitive to wavelength (Figure 1). Based on endogenous contrast, PAT can quantify hemoglobin concentrations *in vivo* with both fine spatial resolution and high sensitivity.

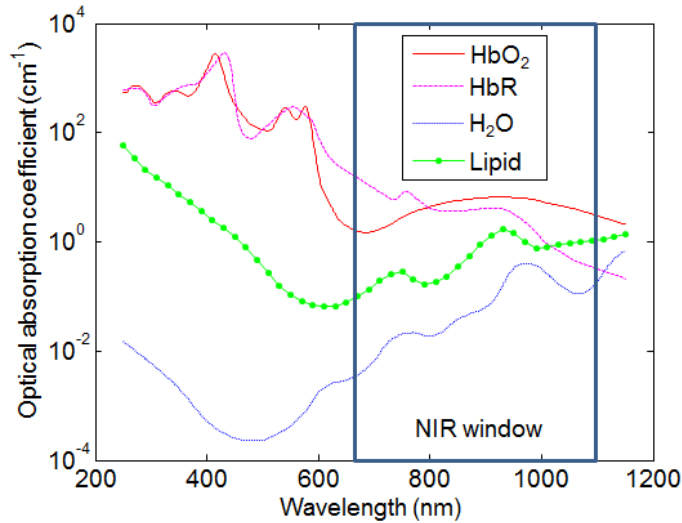

Figure 1. Absorption spectra of oxyhemoglobin ($HbO_2$), deoxyhemoglobin (HbR), water ($H_2O$) and lipid from 250 to 1150 nm.

PAT is based on the photoacoustic effect, and the measurement process works as follows: an optical pulse irradiates an area of the body; then, the pulse energy is partially absorbed by the target and converted into heat, which generates a local transient temperature rise, followed by a local pressure rise. The pressure propagates as ultrasonic waves that are detected by ultrasonic transducers placed outside the body. The image is then formed by mathematically reconstructing the origins of the photoacoustic waves [4, 5]. The photoacoustic signal is proportional to the product of the absorption coefficient $\mu_a$, and the local optical fluence $F$, as expressed in

$$p(\vec{r}, \lambda) \propto \mu_a(\vec{r}, \lambda)F(\vec{r}, \lambda),$$ Eq. (1)

where $p(\vec{r}, \lambda)$ is the initial photoacoustic pressure amplitude at the optical wavelength $\lambda$.

Photoacoustic imaging combines the advantages of optical excitation and acoustic detection, resulting in a hybrid imaging modality with high sensitivity to optical absorption contrast and ultrasonic resolution for deep tissue imaging. This affords PAT several advantages over existing methods. First, PAT benefits from the low acoustic scattering of ultrasonic waves in biological tissue. Ultrasonic scattering is approximately two to three orders of magnitude weaker than optical scattering, which allows PAT to achieve high acoustic resolution at depths far beyond the optical diffusion limit (~1 mm in the skin) [6]. Consequently, PAT has higher spatial resolution in deep tissues than pure optical imaging, which relies on strongly scattered photons for spatial resolution. Second, scattered light presents less difficulty to PAT because any absorbed light, regardless of whether it is scattered is converted into sound. PAT has been shown to penetrate multiple centimeters in tissue, which is sufficient for many clinical applications, such as breast imaging [7]. Third, PAT is inherently compatible with US, thereby enabling dual-modality imaging with complementary contrasts.

Because PAT images spatially map the absorbed optical energy density, PAT can be used to quantify hemoglobin concentrations using the intrinsic optical absorption of hemoglobin. The hemoglobin concentrations are derived by solving the following equation for multiple values of $\lambda$

$$\mu_a(\lambda) = \varepsilon_{HbR}(\lambda)[\text{HbR}] + \varepsilon_{HbO2}(\lambda)[\text{HbO}_2],$$ Eq. (2)

where $\varepsilon_{HbR}$ and $\varepsilon_{HbO2}$ are the known molar extinction coefficients of deoxyhemoglobin (HbR) and oxyhemoglobin (HbO$_2$) at wavelength $\lambda$ [8,9], and [HbR] and [HbO$_2$] are their concentrations respectively. These concentrations can then be used to estimate the oxygen saturation (sO$_2$) defined as

$$\text{sO}_2 = [\text{HbO}_2] / ([\text{HbO}_2] + [\text{HbR}]).$$ Eq. (3)

## 2.2 Combined Photoacoustic and Ultrasound Imaging System

We developed a PAT/US imaging system around a modified clinical US scanner (iU22, Philips Healthcare) as shown in Figure 2. The light sources included a wavelength-tunable dye laser (PrecisionScan-P, Sirah) and a Q-switched Nd:YAG laser (QuantaRay PRO-350-10, Spectra-Physics). The lasers emitted 6.5 ns pulses at a repetition rate of 10 Hz, which were coupled to a fused end, bifurcated fiber bundle that flanked both sides of a linear ultrasound array (L8-4, Philips). The per-channel data was transferred to a custom-built data acquisition (DAQ) system for image reconstruction and display. The DAQ system was synchronized with laser firings by an FPGA-based electronic board that also performed PAT image reconstruction based upon a delay-and-sum beamforming algorithm. The system was capable of displaying PAT, US, and co-registered images live at 5 frames per second. The optical fluence on the skin was less than 10 mJ/cm$^2$, which was within the ANSI safety limit [10].

The system was designed to operate in two different modes: the first mode acquires and displays PAT and US co-registered images (PAT/US mode), while the second mode acquires and displays dual-wavelength photoacoustic images (dual PA mode). During PAT/US mode, the PAT and US images were collected alternately. At the same time, they were processed and overlaid, allowing the use of the US images to accurately determinate the source of the PA signals. The co-registered images were displayed at 5 frames/s. For the dual-PA mode, we selected two wavelengths, 782 nm from the dye laser and 1064 nm from the Nd:YAG laser. A mechanical shutter was synchronized to the laser trigger to alternately block either the dye output or the fundamental 1064 nm of the Nd:YAG laser output on consecutive laser shots. The two wavelengths were routed to the same path by a dichroic mirror before being coupled to the fiber bundle.

## 2.3 Participant Population

The Washington University School of Medicine Institutional Review Board, St. Louis, Missouri, approved this study, and all participants gave informed consents. Eligibility requirements included being 18 or older, healthy, and with the ability to give informed consent. Exclusions included minors, individuals unable to give informed consent, and pregnant women.

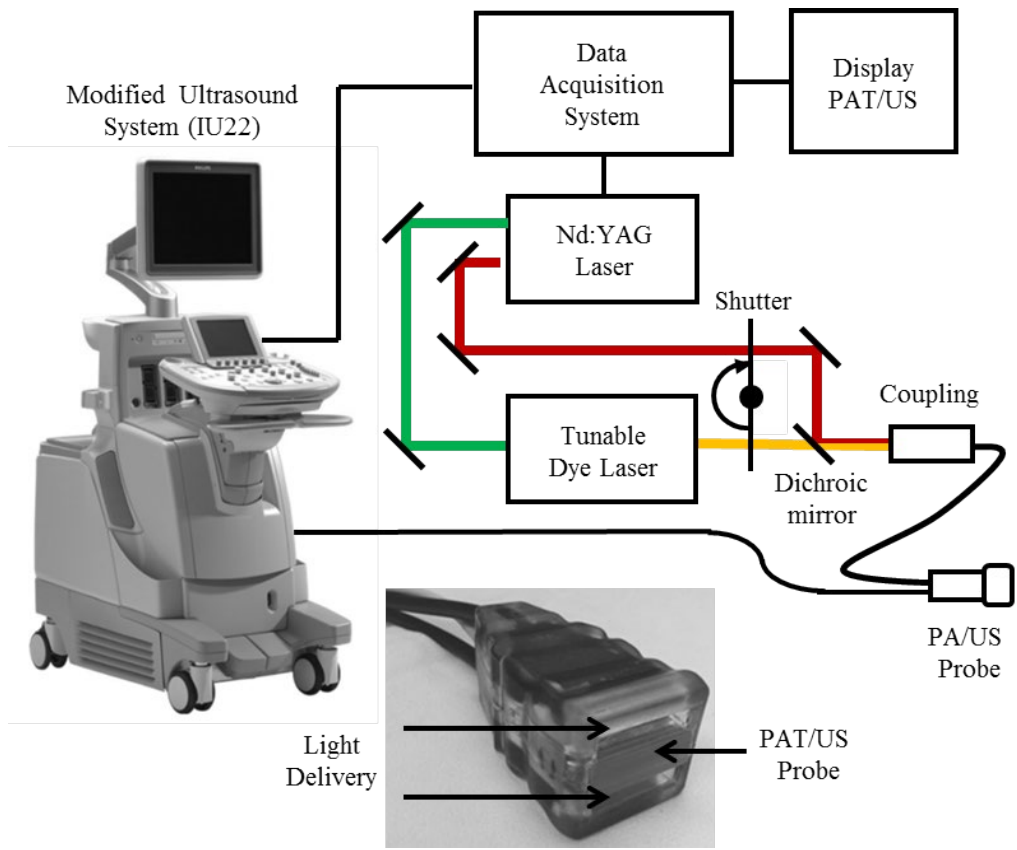

Figure 2. Schematic of the dual-modality photoacoustic and ultrasound imaging system and improved ultrasonic transducer design. The transducer features a flat surface for light delivery and fiber bundles integrated into the transducer housing for easier handling.

*2.4 Procedures*

First, the ultrasound transducer was held perpendicular to the neck and in contact with the skin. US images were used to locate the internal jugular vein. We acquired co-registered PAT and US images and identified the PA signal from the IJV; PAT and US data were collected alternately. After localizing the internal jugular vein, we switched to PA mode and collected PAT images generated at 782 and 1064 nm. The PA signal for the IJV was segmented and averaged. The PAT/US system could estimate oxygen saturation in real time, i.e., every 2 frames (five times per second).

All methods and experimental procedures were carried out in accordance with the approved guidelines of the Institutional Review Board of the Washington University School of Medicine. All experimental protocols were approved by the Institutional Review Board of the Washington University School of Medicine.

## RESULTS

Figure 3 shows PAT, US, and co-registered images of the internal jugular vein acquired from a healthy normal volunteer. The combination of US and PA images allowed us to accurately locate the PA signal from the IJV.

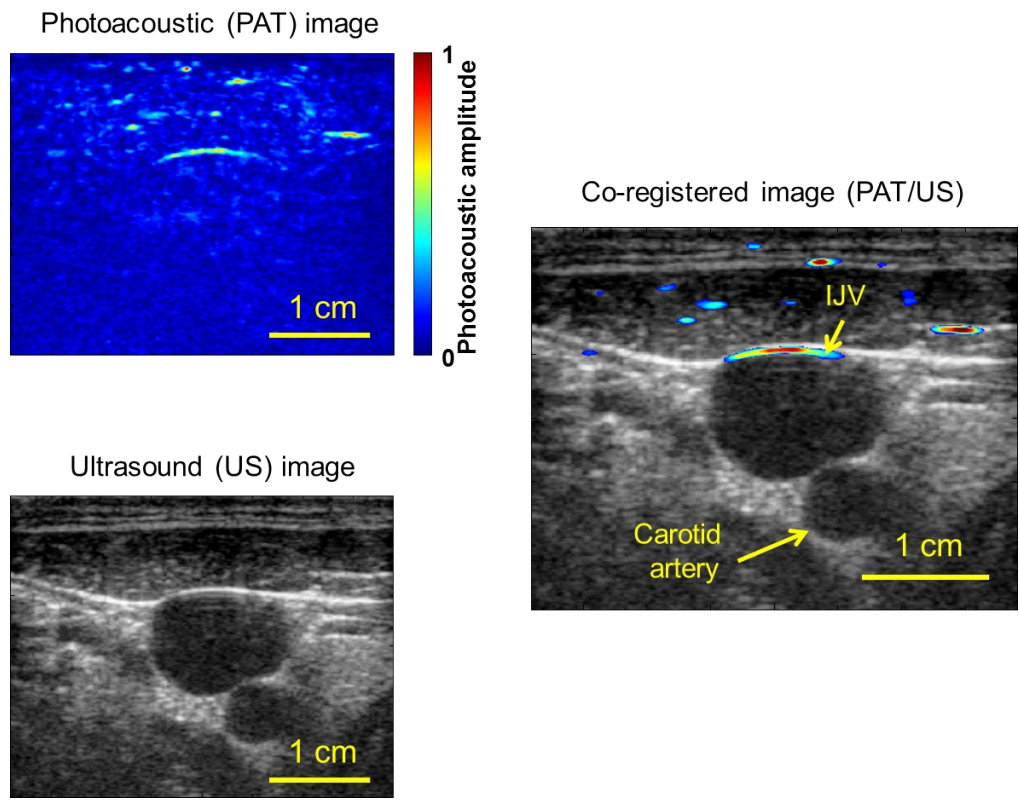

Figure 3. *In vivo* images of the internal jugular vein (IJV) acquired using PAT/US system. (a) *In vivo* PA image (b) *In vivo* US image. (c) Co-registered PAT/US image of the IJV.

During the course of this clinical study the IJV was best visualized by placing the probe along the cross-sectional and longitudinal views of the IJV. The average oxygen saturation was calculated using the conventional PA amplitude method. The average oxygen saturation was 67% for this case, and Table 1 summarizes the results from seven healthy volunteers. Figure 4 shows the simultaneous photoacoustic and ultrasound images of the IJV and the carotid artery (CA). The estimated oxygen saturations in the CA and IJV were 96 and 66%, respectively.

Table 1. Oxygen saturations in internal jugular veins ($s_{ijv}O_2$) of seven healthy volunteers, sensed *in vivo*.

| Volunteer | $s_{ijv}O_2$ (%) |
|---|---|
| 1 | 78 |
| 2 | 67 |
| 3 | 76 |
| 4 | 75 |
| 5 | 79 |
| 6 | 62 |
| 7 | 66 |
| Average | 72 |
| Standard deviation | 7 |
| Normal physiology | 60-75 |

The results demonstrated that the system provides the functional imaging capabilities for quantification of total concentration of hemoglobin (in arbitrary units) and oxygen saturation. These preliminary results are encouraging and agree with those reported in the literature [11]. Since a large fraction (>90%) of internal jugular vein drainage comes from intracranial structures, we expect that changes in global oxygen consumption are mirrored closely by the changes of $s_{ijv}O_2$. Additionally, this concept may also be used to monitor perfusions of other major organs, for example, by measuring venous saturation in the hepatic vein, femoral vein, and renal vein.

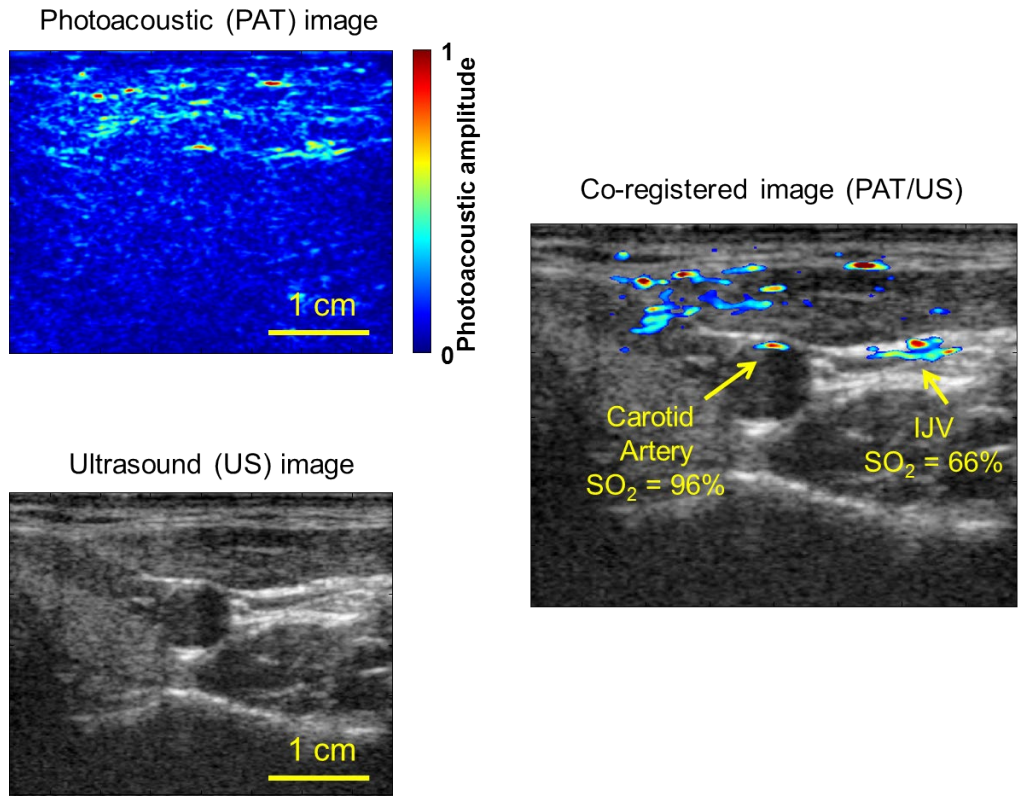

Figure 4. Simultaneous photoacoustic and ultrasound images of the IJV and the carotid artery (CA).

## DISCUSSION

Currently available bedside noninvasive monitors have failed to reliably detect either brief or prolonged episodes of global brain ischemia. Near-infrared spectroscopy (NIRS) is one non-invasive method that is used to monitor the oxygen saturation of a composite tissue located under the probe [12]. NIRS can evaluate the relationship between oxygen availability and oxygen consumption. However, a major limitation of NIRS monitoring is their key assumption that the cerebral blood in the path of the near-infrared light is composed of 15-20% arterial, 70-80% venous, and 5-10% capillary blood. This assumption may not hold true in many patients, especially in children [13,14].

NIRS takes advantage of the near-infrared window in biological tissue (650-1100 nm) to improve light penetration into biologic material such as skin, fat, and muscle (Figure 1) [15, 16]. Current NIRS monitors are hampered by a number of limitations. First, the inability to calibrate the depth of measurements of regional oxygen saturation ($sO_2$) can lead to inaccurate estimates, due to extra-cranial contamination [17]. Second, there can be considerable biological variation in individual cerebral arterial/venous/capillary ratios between patients. Third, variation among individuals in skull thickness and cerebrospinal fluid volume can significantly alter tissue oxygen saturation values [18].

We successfully developed a dual-wavelength photoacoustic and integrated ultrasound system capable of real-time noninvasive measurement of oxygen saturation in the internal jugular vein. These preliminary results raise confidence that this emerging technology can be used clinically for accurate monitoring of $s_{ijv}O_2$. We expect that changes in the global oxygen consumption of the brain can be mirrored closely by changes in $s_{ijv}O_2$. The results from this study are an important step towards clinical translation of photoacoustic imaging into a noninvasive tool for monitoring $s_{ijv}O_2$ to prevent brain hypoxic-ischemic injuries. In addition, fast co-registration of PAT and US images makes the technique suitable for studying the temporal variations of oxygen saturation in response to physiologic challenges in clinical settings.

## ACKNOWLEDGEMENTS

We thank Prof. James Ballard for his attentive reading of the manuscript. This work was sponsored by NIH grants U54 CA136398, R01 CA134539, DP1 EB016986 (NIH Director's Pioneer Award), and R01 CA186567 (NIH Director's Transformative Research Award).

## DECLARATION OF INTEREST

L. V. Wang has a financial interest in Microphotoacoustics, Inc., CalPACT, LLC, and Union Photoacoustic Technologies, Ltd., which, however, did not support this work.